 \documentclass[final,1p,times,twocolumn]{elsarticle}

\usepackage{amssymb}
\usepackage{url}

\newcommand\op[1]{\mathop{\rm #1}\nolimits}

\newcommand\N{{\mathbb N}}

\newcommand\R{{\mathbb R}}

\newcommand{\weg}[1]{}

\usepackage{fullpage}
\usepackage{color}
\journal{Physica A}

\begin{document}

\begin{frontmatter}


%

 \title{Modeling electricity spot prices using mean-reverting multifractal processes}


\author{Martin Rypdal\corref{cor1}}
 \ead{martin.rypdal@uit.no}
\author{Ola L{\o}vsletten}
\address{Department of Mathematics and Statistics, University of Troms{\o}, N-9037 Troms{\o}, Norway.}
 \cortext[cor1]{Corresponding author}

\begin{abstract}
We discuss stochastic modeling of volatility persistence and anti-correlations in electricity spot prices, and for this purpose we present two mean-reverting versions of the multifractal random walk (MRW). In the first model the anti-correlations are modeled in the same way as in an Ornstein-Uhlenbeck process, i.e. via a drift (damping) term, and in the second model the anti-correlations are included by letting the innovations in the MRW model be fractional Gaussian noise with $H<1/2$. For both  models we present approximate maximum likelihood methods, and we apply these methods to estimate the parameters for the spot prices in the Nordic electricity market. The maximum likelihood estimates show that electricity spot prices are characterized by scaling exponents that are significantly different from the corresponding exponents in stock markets, confirming the exceptional nature of the electricity market. In order to compare the damped MRW model with the fractional MRW model we use ensemble simulations and wavelet-based variograms, and we observe that certain features of 
the spot prices are better described by the damped  MRW model. The characteristic correlation time is estimated to approximately half a year.  
\end{abstract}

\end{frontmatter}


\section{Introduction} \label{introduction}
Since the 1990s, several of the world's electricity markets have been deregulated and re-organized in order to introduce competition and increase efficiency \cite{Bye:2005ww}. In the de-regularized electricity markets there is trading of contracts for physical delivery of electric energy at a certain hour the next day. The price of such a contract is called the electricity spot price, and it is recorded for every hour of the year. The records of historical spot prices are extremely interesting from a scientific point of view, and a lot of effort has been devoted to describing and modeling their dynamics. 

In this paper we analyze data from the Nordic electricity spot market (Nord Pool), which
is known to exhibit a daily periodicity, a weekly periodicity and a one-year periodicity. These periodicities can be understood from a simple analysis of supply and demand. The consumption of electric energy is generally lower at night than during the day, and this causes a daily cycle in price. In the same way, the industry's demand for electric energy is lower during the weekend, and in the Nordic countries the demand for electric energy increases in winter due to the need for heating. In addition, the Nord Pool market is largely based on hydroelectric energy which supply has a seasonal variation. 

On top of the periodic variations, the electricity spot prices have more unpredictable changes which are related to factors such as the weather, the distribution network, the level of industrial activity and general market dynamics. 
The aim of this paper is to present models for these non-periodic fluctuations. This task is important for several reasons. For instance, accurate quantification of the variability of spot prices is essential for correct  pricing of futures and other electricity-based derivatives. 
It is also interesting to compare the characteristics of electricity spot prices with the prices of other commodities. It is observed that some of the ``stylized facts'' of electricity spot prices are similar to what is seen for stock and currency, whereas other properties are quite different.   

Among the ``stylized facts'' of spot prices that {\em are} similar to other financial time series, are
non-Gaussian distributions of log-returns and long-range volatility dependence. In stock markets it is known that volatility is closely tied to the trading volume \cite{Lobato:2000wr}, and it is reasonable to assume that the same is true for the electricity market. However, since electric energy is expensive to store, the delivered volume must equal the consumption. It is therefore somewhat surprising that electricity spot prices have such clear memory effects in volatility, and it shows that volatility clustering can be present even in markets with limited room for speculative behavior.   
On the other hand, the (non-periodc component of the) demand for electric energy is obviously not constant. It depends on a range of physical and economic factors, which certainly may contain long-range memory effects.

The ``stylized facts'' mentioned above can be described in a parsimonious way using multifractal models, and it has already been suggested by some authors \cite{Norouzzadeh:2007bf,Malo:2006ti,Malo:2009fo} to apply multifractal modeling to electricity spot prices. Within this framework the logarithmic returns are modeled as $x_t=X(t+\Delta t)-X(t)$, where $X(t)$ is a multifractal process with stationary increments. Multifractality means that $X(t)$ is characterized by power-law scaling of its moments, i.e. $E|X(t)|^q \sim t^{\zeta(q)}$, in some range $0<t<T$ or asymptotically as $t \to 0$. For selfsimilar processes, such as Brownian motion and fractional Brownian motion, the scaling function $\zeta(q)$ is linear and its slope equals the selfsimilarity exponent. However in general, the scaling functions are concave and in the following we will refer to multifractal processes as those with strictly concave scaling functions. 

The stochastic modeling presented in this paper is based on the (log-normal) MRW model. This model was introduced by Bacry {\em et al.} \cite{Bacry:2001gl}, and it is often preferred over other multifractal models because of its simplicity and its desirable theoretical properties. For the purpose of modeling financial time series, the most important property of the MRW model (and other multifractal models) is that, even if the log-returns $x_t$ are uncorrelated, the auto-correlation functions for their amptitudes $|x_t|$ decay as power-laws as functions of the time lag $\tau$: 
\begin{equation} \label{volclust}
R_{|x|}(\tau) \sim \tau^{- \lambda^2/4}\,.
\end{equation}
Here $\lambda$ is called the intermittency parameter. This property allows us to model volatility clustering without imposing any particular type of correlations between the returns themselves. Simultaneously, the concave shape of the scaling function $\zeta(q)$ implies that kurtosis of $X(t)$ increases as $t \to 0$. This means that the return distributions are increasingly leptokurtic on shorter time scales, and consequently non-Gaussian. 

In addition to ``fat-tailed'' distributions and slowly decaying volatility dependence, electricity spot prices have anti-correlated returns. 
This was first discovered by Weron \cite{Weron:2000hla} and has since been confirmed by several authors. Anti-correlations are atypical in financial time series and would normally lead to arbitrage possibilities. However, since electricity is expensive to store, such arbitrage possibilities are hard to exploit, and hence anti-correlated returns may exist. There are mainly two ways in which these anti-correlations are modeled: The simplest approach is to consider models similar to Ornstein-Uhlenbeck (OU) processes. The standard OU processes are defined via stochastic differential equations on the form 
\begin{equation} \label{OU}
dX(t) = -\nu\, \big{(}X(t)-m\big{)}\,dt + \sigma\,dB(t)\,,
\end{equation}
where $B(t)$ is a Brownian motion. The first term on the right-hand side of equation (\ref{OU}) is called the drift term (or the damping term), and for $\nu>0$ this causes anti-correlations since it prevents $X(t)$ to diffuse far from its mean value $m$. One can choose initial conditions for $X(t)$ such that the OU process is stationary, and in this case the auto-correlation function for the returns $x_t=X(t+\Delta t)-X(t)$ has an exponential decay with a characteristic time scale $1/\nu$:
\begin{equation} \label{exp}
R_x(\tau)  \sim - \nu^2 e^{-\nu \tau}\,.
\end{equation}
To include the effects of non-Gaussian log-returns and volatility clustering, the white noise $dB(t)$ can be replaced by other types of (uncorrelated) noise processes, such as jump processes with regime switching or (as in this paper) multifractal processes. Examples of non-Gaussian OU-type models for spot prices are given by Benth {\em et al.} \cite{Benth:2007gy}, Erlwein {\em et al.} \cite{Erlwein:2010fn} and Weron {\em et al.} \cite{Weron:2003un}. 

An alternative to including mean reversal via a drift (damping) term is to describe the anti-correlations using Hurst exponents. 
This assumes that the logarithm (of the non-periodic component) of the spot price can be described by a process $X(t)$ with stationary increments and power-law scaling of the variogram. The latter means that the second moment of the differences $\delta X_\tau(t)=X(t+\tau)-X(t)$ is a power-law as a function of the lag $\tau$, in which case $H$ is defined by
\begin{equation} \label{Hdef}
\mathbb{E} [\delta X_\tau(t)^2] \propto \tau^{2H}\,.
\end{equation}
For self-similar processes such as fractional Brownian motions, the Hurst exponents are equal to the selfsimilarity exponents, but in general we do not need to assume selfsimilarity to use Hurst exponents. If the Hurst exponent is well-defined, $H \neq 1/2$ and $X(t)$ has stationary increments, then the auto-correlation function of the log-returns $x_t$ decays as 
\begin{equation} \label{acf1}
R_x(\tau) \sim 2H(2H-1)\,\tau^{2H-2}\,.
\end{equation}
We see that the case $H<1/2$ corresponds to algebraically decaying anti-correlations.  

There exist several methods for estimating Hurst exponents, and various authors have reported different estimates in different electricity markets using different techniques. However, all reported values of $H$ are in the interval $H < 1/2$.  
Weron and Przybylowicz \cite{Weron:2000bj} applied R/S analysis to daily averaged prices form the California Power Exchange (CalPX), and reported a Hurst exponent  around $H=0.42$. Using a wavelet-based estimator Simonsen \cite{Simonsen:2003gw} has estimated a Hurst exponent $H=0.41$ for the Nord Pool market. Anti-persistence in the Nord Pool data is also found by Erzgr{\"a}ber {\em et al.} \cite{Erzgraber:2008kn} by R/S-analysis. Using de-trended fluctuation analysis Norouzzadeh {\em et al.} \cite{Norouzzadeh:2007bf} have estimated the Hurst exponent of the Spanish electricity exchange,  Compania O Peradora del Mercado de Electricidad (OMEL), to be $H=0.16$. 

In this work we present two stochastic models for electricity spot prices. In the first model, which we will refer to as a damped MRW model, we consider a process of OU type, but where we have introduced stochastic volatility in the same way as in a standard MRW model. This type of model (which was first introduced in \cite{Rypdal:2011gt} to describe magnetic field fluctuations in the turbulent solar wind) has exponentially decaying anti-correlation of returns and algebraically decaying dependence in volatility. 

The second model is referred to as a fractional MRW process. Here we use the same stochastic volatility as in the standard MRW model, but instead of a white noise, the process is driven by a fractional Gaussian noise with Hurst exponent $H<1/2$. 

The main results of this paper are presented in sections \ref{modeling} and \ref{ML}. In section \ref{modeling} we discuss methods for modeling spot prices using mean-reverting multifractal processes, and in section \ref{ML} we derive approximate ML estimators. 
These methods are generalizations of a recently developed method for inference on standard (uncorrelated) MRW models \cite{Lovsletten:2011vm}. In section \ref{results} we apply the ML estimators to the Nord Pool data and show that the estimates are consistent with the preliminary analysis presented in section \ref{data}. 

As a part of our further analysis we compare our two models in order to determine which provides the better description of the Nord Pool data. To do this we construct ensembles of synthetic signals from the two models, with periodic components added, and compute wavelet-based variograms. These variograms are then compared to the corresponding variogram for the Nord Pool data. The details and results of this test are described in section \ref{testing}. In section \ref{conclusion} we give some concluding remarks. 

\noindent
\begin{figure}[t]
\begin{center}
\includegraphics[width=15.0cm]{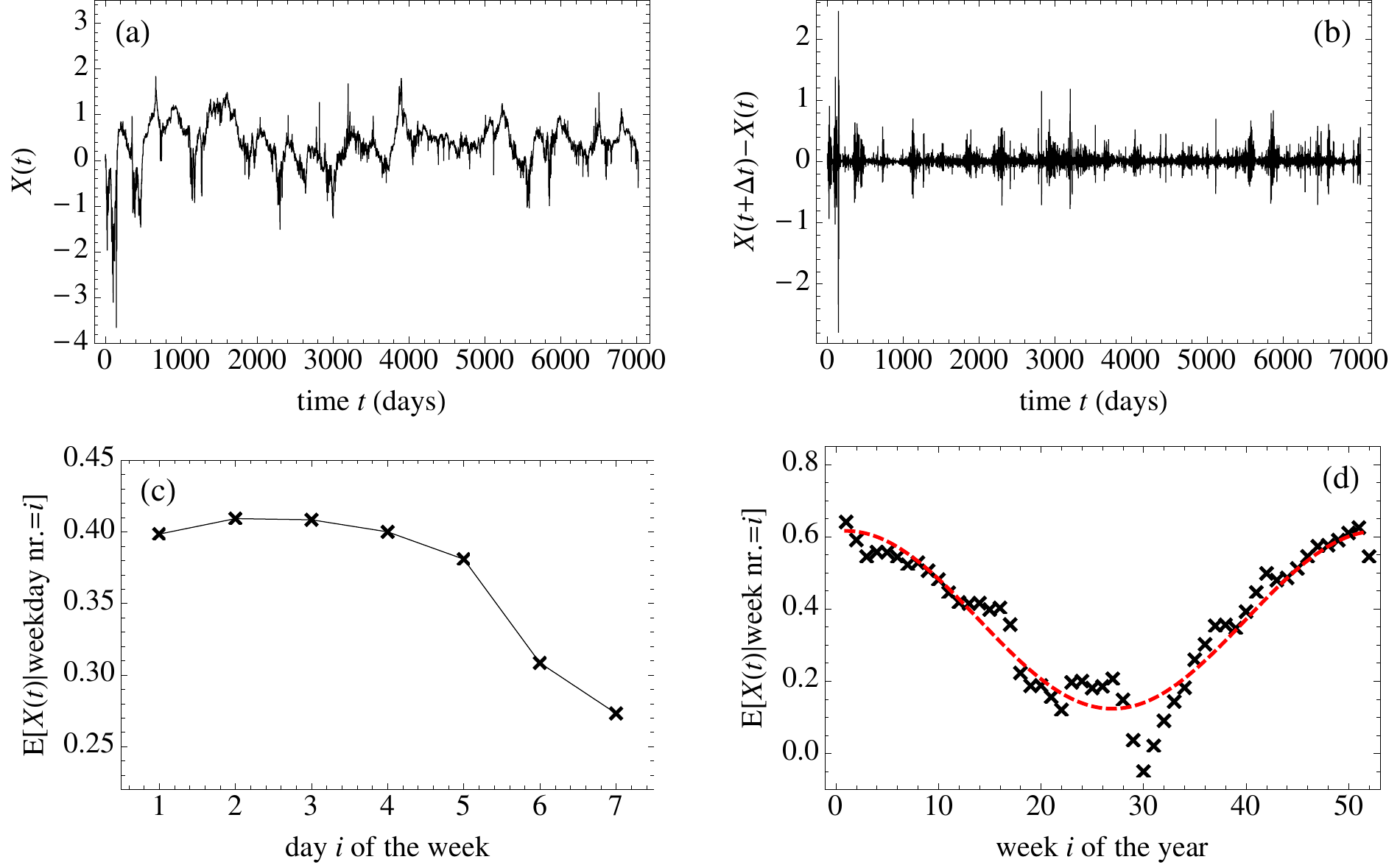}
\caption{(a): The time series obtained from the spot prices by taking the logarithm and subtracting a linear trend. The plotted signal consists of daily means from May 4th 1992 to August 27th 2011. (b): The increments of the time series plotted in (a). (c%
): For the signal in (a) we plot the mean value conditioned on the weekday. This means that the first point is is obtained from the signal in (a) by taking the mean value over all Mondays, the second is obtained by taking the mean over all Tuesdays, and so on.  (d): The weekly means of the signal in (a) conditioned on the week of the year. The dotted line shows a fitted sinusoidal oscillation with a period of one year and an amplitude $0.25$. 
} \label{fig1}
\end{center}
\end{figure}
\noindent
\begin{figure}[t]
\begin{center}
\includegraphics[width=13.0cm]{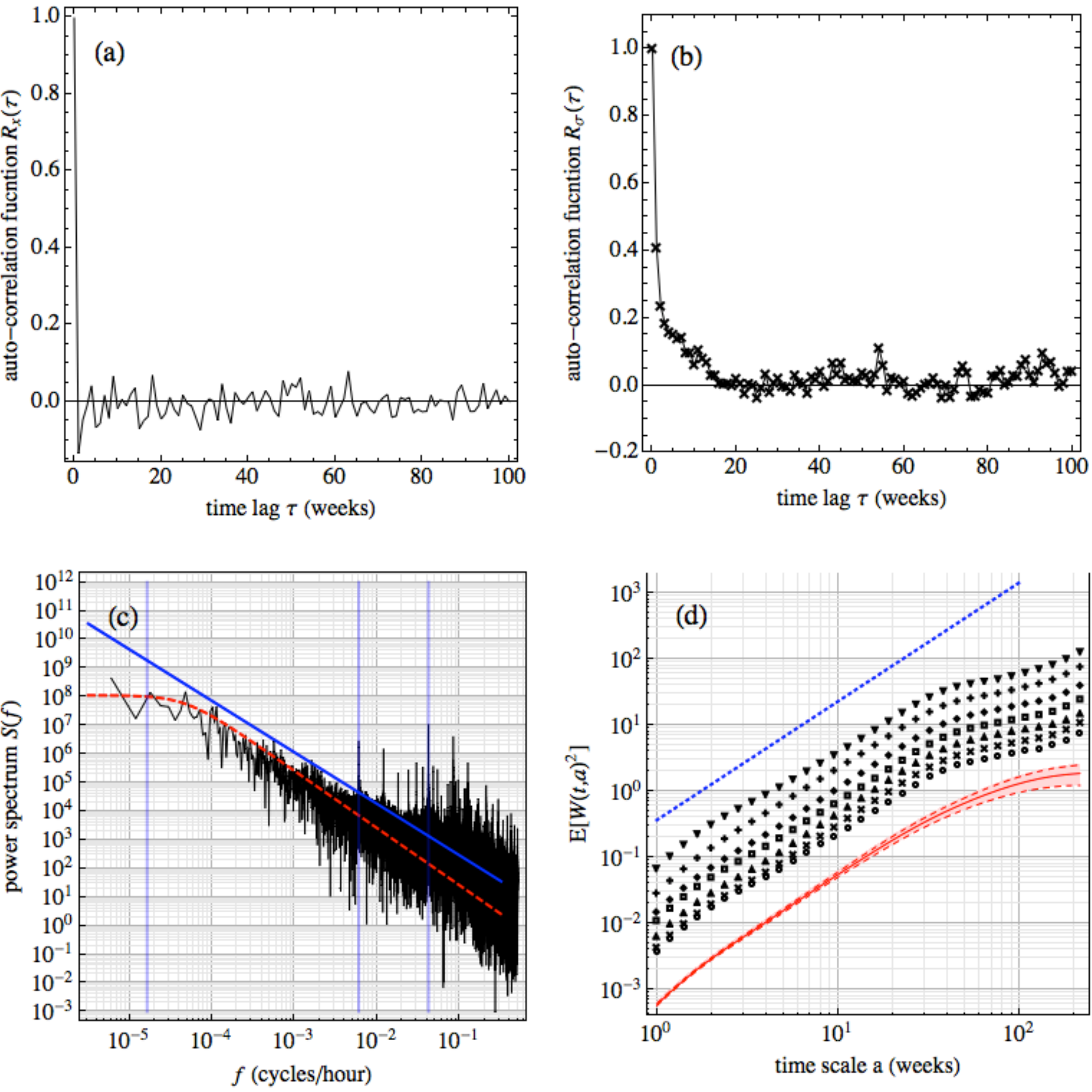}
\caption{(a): The auto-correlation function  of the daily mean log-returns sampled at weekly intervals on a given weekday, and then averaged over the seven weekdays. (b): The auto-correlation function for the absolute values of the daily mean log-returns sampled on Mondays. (c%
): Double logarithmic plot of the power spectrum of the daily mean spot prices. The vertical lines correspond to periods of one year, one week and one day. The solid line represents a power law $1/f^{2H+1}$ with $H=0.4$, whereas the dotted line is a Lorentzian spectrum $\nu/(\nu^2 + (2 \pi f)^2)$, with $1/\nu = 20 \, \mbox{weeks}$. (d) Double logarithmic plots of the wavelet-based variograms for the log-returns sampled on different weekdays. The curves are shifted to make them all visible (Mondays through Sundays are sorted from bottom to top.) The dotted line is a power law corresponding $H=0.4$. The bottom curves show the results of the wavelet-based variograms estimated from realizations of a OU process with parameters $1/\nu=20\,\mbox{weeks}$. For each time scale $a$ we have plotted the mean (solid curve) and the $1/8$ (upper and lower) quantiles (dashed curves).} \label{prelimfig}
\end{center}
\end{figure}
\section{Description of data and preliminary analysis} \label{data}
The data analyzed in this paper are received from the Data Administrator at Nord Pool Spot (\url{http://www.nordpoolspot.com/}) upon request. The data set consists of hourly spot prices measured in Norwegian Kroner (NOK) from May 4th 1992 to August 27th 2011.   
The one-day period and the seven-day period show up as peaks in the estimated power spectrum, as do their higher harmonics. This can be seen in figure \ref{prelimfig}(c).

To eliminate the effect of the strong daily periodicity we consider the signal $P(t)$ of daily mean prices, and we denote $X(t)=-\mu\, t + \log P(t)$, where $\mu$ is the mean daily log-return, i.e. 
$$
\mu = \mathbb{E}\Big{[}\log \frac{P(t)}{P(t-\Delta t)}\Big{]}\,,
$$ 
with $\Delta t = 1\,\mbox{day}$. The resulting signal $X(t)$ is shown in figure \ref{fig1}(a). 
We will refer to the time series $X(t+\Delta t)-X(t)$ as the daily log-returns, and they are shown if figure \ref{fig1}(b).  
The sample standard deviation of this signal is $\sigma=0.12$.  

Even if we study daily mean values, there are still seven-day and one-year periodicities in the signal. 
This is illustrated in figures \ref{fig1}(c%
) and \ref{fig1}(d). In figure \ref{fig1}(c%
) we have plotted the conditional mean of $X(t)$ given that the signal is sampled only on a specific day of the week. We see from the figure that the spot price is lower for Saturdays and Sundays. The conditional mean of $X(t)$ has a variation of about $0.15$ from weekdays to the weekend. This variation is rather large, roughly equal to the standard deviation of the log-returns, and therefore the seven-day periodicity exerts  substantial influence on certain estimates. For instance, the estimated auto-correlation function of the daily log-returns has a strong seven-day period which makes it hard to analyze how the correlations decay in the non-periodic component of the signal. 

Due to the strong weekly periodicity we will for most of the remaining analysis consider a discretization $y_k = X(k \Delta t)$, where $\Delta t = 1 \,\mbox{week}$. This simply means that we consider the signal $X(t)$ sampled once a week, for instance every Monday. This gives us seven (dependent) time series, and as we will see in section \ref{results}, the parameter estimates vary little between these time series, with some exceptions for Saturdays and Sundays.         

In the weekly sampled signals there is still a one-year periodicity. This is illustrated in figure \ref{fig1}(d), where we have plotted the weekly mean of $X(t)$ conditioned on the week number of the year. The seasonal dependence can very roughly be described by a sinusoidal oscillation with amplitude $0.25$. This amplitude is greater than the the weekly oscillation, but since the period is about 50 times greater, the one-year oscillation exerts much less influence on the analysis. We make no attempt to de-trend this seasonality, but  we have crudely tested the effect that a sinusoidal oscillation with amplitude of $0.25$ has on the estimates that we perform. For the ML estimators presented in section \ref{ML} we observe that the addition of the sinusoidal signal slightly decreases the estimate of $H$ in the fractional MRW model, and slightly increases the time scale $1/\nu$ in the damped MRW model. The relative changes in $H$ and $1/\nu$ are typically $<10\%$, whereas the intermittency parameter is almost unchanged. Other estimates, such as the auto-correlation function for the fluctuation amplitudes $|X(t+\Delta t)-X(t)|$, are more influenced by the seasonal variation. We have plotted this correlation function in figure \ref{prelimfig}(b), and we clearly see a one-year oscillation on top of the decay.  
Simonsen \cite{Simonsen:2005dn} has found that this correlation function decays algebraically as $1/\tau^{0.07}$, and if we compare with equation (\ref{volclust}), this corresponds to $\lambda = 0.53$. This value of the intermittency parameter is slightly higher than what is estimated for stock markets, which typically are in the range 0.3-0.4 \cite{Lovsletten:2011vm,Bacry:2008by}. This difference between stock markets and electricity spot markets is confirmed by the ML estimates presented in section \ref{results}.   
 
 The estimated correlation functions for the log-returns themselves are plotted in figure \ref{prelimfig}(a). It is the average of the correlation functions for the seven time series  obtained by sampling with weekly intervals on a given weekday. It might be possible to detect some anti-correlation from this figure, but it is impossible to distinguish between an exponential and an algebraic decay, i.e. between the expressions in equations (\ref{exp}) and (\ref{acf1}). The correlations of returns are better analyzed using a wavelet-based variogram 
 $ V(a) = \mathbb{E}[|W(t,a)|^2] $, where  
\begin{equation} \label{wavelet}
 W(t,a) = \frac{1}{\sqrt{a}} \int X(t') \, \psi \Big{(} \frac{t'-t}{a}\Big{)}\,dt'\,, 
 \end{equation}
is the wavelet transform of $X(t)$ with respect to the mother wavelet $\psi$. The wavelet transform scales as \cite{Muzy:1991vm}: 
\begin{equation} \label{skalering1}
W(t,a) \sim \sqrt{a} \,\Big{(} X(t+a)-X(t) \Big{)}\,,
\end{equation}
and hence, if the Hurst exponent of $X(t)$ is well defined\footnote{The Hurst exponent is well defined if equation (\ref{Hdef}) holds, and we do not need to assume that $X(t)$ is selfsimilar nor multifractal. Equation (\ref{skalering2}) follows from (\ref{skalering1}) by the definition of $H$.} we have 
\begin{equation} \label{skalering2}
V(a) \sim a^{2H+1}\,.
\end{equation}

In figure \ref{prelimfig}(d) we show the wavelet-based variograms for the weekly sampled data estimated using a wavelet $\psi$ that is the first derivative of a Gaussian. 
The dotted line above the variograms has slope $1.8$ in the double-logarithmic plot, corresponding to a Hurst exponent $H=0.4$. A striking feature in figure \ref{prelimfig}(d) is the sharp 
``breaks'' in the variograms in the range 20-50 weeks. This represents a characteristic time scale, and the existence of characteristic scales is not consistent with an algebraic decay of the auto-correlation function. In fact, this feature of the spot price signal 
suggests a process of OU-type, for which the correlation decay has a characteristic time scale $1/\nu$. To support this statement we have simulated an ensemble of OU processes with $1/\nu= 20 \,\mbox{weeks}$ and estimate the wavelet-based variograms using the same method as for the spot-price data. The results of this analysis are shown as the bottom curves in figure \ref{prelimfig}(d). We see from these plots that the OU-process captures the flattening of the variogram on long time scales. The same feature is seen in the power spectrum, which is plotted in figure \ref{prelimfig}(c%
).  As opposed to the pure power-law spectrum, a Lorentzian spectrum, which scales as $\sim 1/f^2$ for $f \gg \nu$, captures the flattening of the spectrum at low frequencies.  This seems to indicate that a model of OU type is preferable over a fractional model. However, we must take into account that the observation of a characteristic time scale may be an effect of the one-year periodicity, and that a scale-invariant description of the non-periodic component  may nevertheless be appropriate. These questions are discussed in more detail in section \ref{testing}.

\section{Modeling anti-correlations and intermittency} \label{modeling}
In this section we present two versions of the MRW model. We begin by giving a brief description of the standard MRW model.

\subsection{Stochastic volatility and MRW processes} \label{MRWsection}
Consider a discretization of the logarithmic price $X(t)$, i.e. $y_k = X(k \Delta t)$. This signal is modeled as 
\begin{equation} \label{mrw1}
y_k =  y_{k-1} + \sigma\,\sqrt{M_k} \,\varepsilon_k + \mu\,\Delta t\,,
\end{equation}
where $\varepsilon_k$ is a standard Gaussian white noise. The stochastic volatility term is defined as $M_k = c\,\exp(h_k)$, where $h_k$ is a centered Gaussian process with co-variances 
\begin{equation} \label{mrw2}
\op{Cov}(h_k,h_l) = \lambda^2 \log^+ \frac{T}{(|k-l|+1)\Delta t}\,.
\end{equation}
The constant $c$ is chosen so that $1/c=\mathbb{E}[\exp(h_k)]$. Note that for $\lambda=0$ the logarithmic price is a Brownian motion with drift, i.e. the price is modeled as a geometric Brownian motion.

From equations (\ref{mrw1}) and (\ref{mrw2}) one can derive equation (\ref{volclust}), which shows that the MRW model describes algebraically decaying volatility dependence. 
This type of long-range dependence is common in financial time series, and the MRW model has shown to provide good descriptions of the fluctuations of stock prices and currency exchange rates \cite{Bacry:2008by}. However, the (drift-compensated) log-returns $x_k = y_k-y_{k-1}-\mu\, \Delta t$ are uncorrelated in this model, and therefore the model needs to be modified to make it capture the anti-correlations of spot-price data.

\subsection{A dampled MRW model}
The discrete-time analog of OU processes are auto-regressive models of order one (AR(1) processes). These can be written on the form
\begin{equation} \label{damped}
y_k = \phi y_{k-1} +\sigma\,\varepsilon_k + \mu\, \Delta t\,,
\end{equation}
where $\phi=1-\nu\,\Delta t$. This model can now be generalized to include multifractal volatility by replacing the constant $\sigma$ with the process $\sigma\,\sqrt{M_k}$, where $M_k = c\,\exp(h_k)$ is as defined above. The resulting model is given by the following equation 
\begin{equation} \label{damped}
y_k = \phi y_{k-1} + \sigma\,\sqrt{M_k}\,\varepsilon_k\,.
\end{equation}
We will refer to this process as a damped MRW model. We have here assumed that $\mu=0$.\footnote{This can be done without loss of generality since the parameter $\mu$ is easily estimated from data. One can then replace the logarithmic prices $X(t)$ with the drift-compensated signal $X(t)-\mu t$.}

Keeping in mind that $M_k$ and $\varepsilon_k$ are independent processes, and that $\sqrt{M_k}$ is normalized to have unit variance, we can 
use equation (\ref{damped}) to derive some simple properties of the damped MRW model. 
The main observation is that the stochastic volatility term $\sqrt{M_k}$ does not effect the second order statistics. This means that the auto-correlation functions, variograms and power spectra are the same for the damped MRW process as for the corresponding AR(1) process. For instance, for $k>0$, the auto-correlation function is
$$
\mathbb{E}[y_0y_k] =\frac{\sigma^2}{1-\phi^2}\,\phi^{k} \, \stackrel{\Delta t \ll 1/\nu}{\approx} \, \frac{\sigma^2}{1-\phi^2}\,e^{-\nu\, t}\,,
$$
where $t = k \Delta t$. The auto-correlation function for the log-returns $x_k=y_k-y_{k-1}$ is then 
\begin{equation} \label{dampedacf}
\mathbb{E}[x_0x_k]  \sim - \Delta t^2 \frac{d^2}{dt^2} \frac{\sigma^2}{1-\phi^2}\,e^{-\nu\, t}
=  -\frac{\Delta t^2 \sigma^2 \nu^2}{1-\phi^2}\,e^{-\nu\, t}
\,,
\end{equation}
and (up to discreteness effects) the power spectrum of $y_k$ is a Lorentzian: 
$$
S(f) \propto \frac{\nu}{\nu^2 + (2 \pi f)^2}\,.
$$

Although the damped MRW models and OU processes (or AR(1) processes) share the same correlations, there are essential differences between the models. As explained in the introduction and in section \ref{MRWsection}, the factor $\sqrt{M_k}$  introduces volatility clustering and ``fat tailed'' return distributions, features that are not contained in standard OU processes. This can be seen from figure \ref{examplefig}. In figures \ref{examplefig}(a) and \ref{examplefig}(b) we show a realization of an OU process $X(t)$ and its increments $X(t+\Delta t)-X(t)$ respectively, and in figures \ref{examplefig}(e) and \ref{examplefig}(f) we show the corresponding plots for a damped MRW process. We clearly see that the increments of the damped MRW model has volatility clustering and spikiness that is not present in the OU process. If we compare with the Nord Pool data, which are shown in figures \ref{examplefig}(i) and \ref{examplefig}(j), we observe that these features are essential for accurate modeling of spot prices.

\subsection{A fractional MRW model}

The second class of models that we introduce are fractional MRW models. Here the term ``fractional'' refers to the replacement of the white Gaussian noise $\varepsilon_k$ with a fractional Gaussian noise 
$$
\varepsilon_k^{(H)} = B_H(k+1)-B_H(k)\,.
$$
Here $B_H(\cdot)$ is a fractional Brownian motion with selfsimilarity exponent $H$.  
This gives processes on the form
\begin{equation} \label{model1}
y_k = y_{k-1} + \sqrt{M_k}\,\varepsilon^{(H)}_k\,,
\end{equation}
where $M_k$ is as described in section \ref{MRWsection}. 

The parameter $\phi$ is absent from this model since the anti-correlations of returns are described via a Hurst exponent $H<1/2$. In fact, for $H \neq 1/2$, the correlations of $x_k = y_k-y_{k-1}$ are given by the expression
\begin{equation} \label{fractionalacf}
\mathbb{E}[x_0x_k]  \sim 2H(2H-1) \,k^{2H-2-\lambda^2/4}\,.
\end{equation}
If we compare equation (\ref{fractionalacf}) with equation (\ref{dampedacf}) we see that a distinguishing feature for the two models is that the auto-correlation function for log-returns decays exponentially for the damped MRW model, whereas it decays algebraically for the fractional MRW model. 

\begin{figure}[t]
\begin{center}
\includegraphics[width=15.0cm]{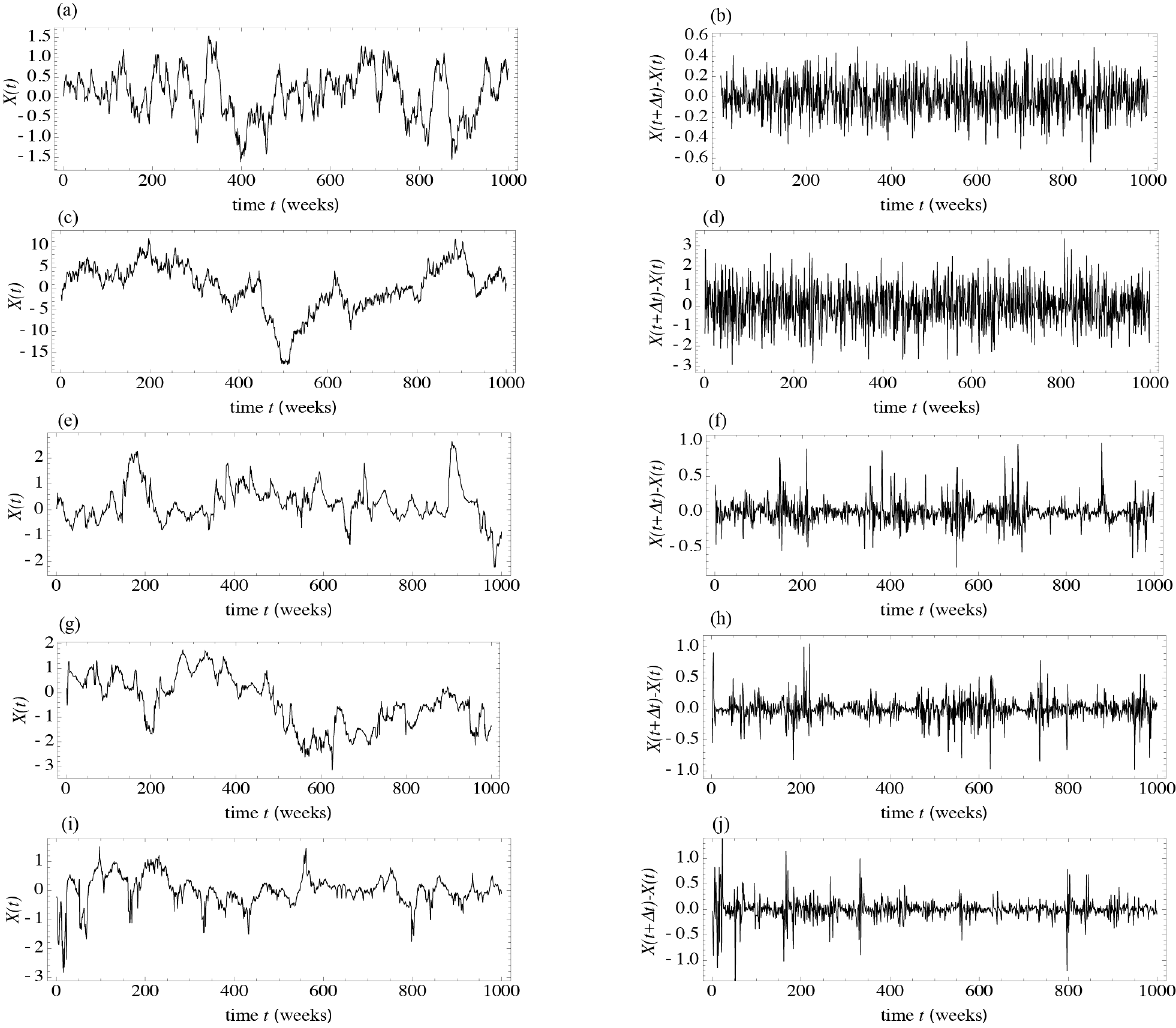}
\caption{(a): A realization of an OU process with $1/\nu= 20 \,\mbox{weeks}$. (b): The increments of the signal in (a). (c%
): A realization of a fractional Brownian motion with $H=0.4$. (d): The increments of the signal in (c%
). (e): A realization of a damped MRW model with $1/\nu= 20 \,\mbox{weeks}$ and $\lambda=0.7$. (f): The increments of the signal in (e). 
(g): A realization of a fractional MRW model with $H=0.4$ and $\lambda=0.7$. (h): The increments of the signal in (g). 
(i): The daily mean (logarithmic) spot price sampled every Monday. (j): The increments of the signal in (i).  
} \label{examplefig}
\end{center}
\end{figure}

As for the damped MRW model, the factor $\sqrt{M_k}$ in introduces volatility clustering and ``fat tailed'' return distributions.  
In figures \ref{examplefig}(c) and \ref{examplefig}(d) we show a realization of a fractional Brownian motion $X(t)$ with $H=0.4$ and its increments $X(t+\Delta t)-X(t)$ respectively. In figures \ref{examplefig}(g) and \ref{examplefig}(h) we show the corresponding plots for a fractional MRW process with $H=0.4$. We see that the increments of the fractional MRW model has volatility clustering and spikiness that is not present in the fractional Brownian motion.

\noindent
\noindent
\begin{figure}[t]
\begin{center}
\includegraphics[width=15.0cm]{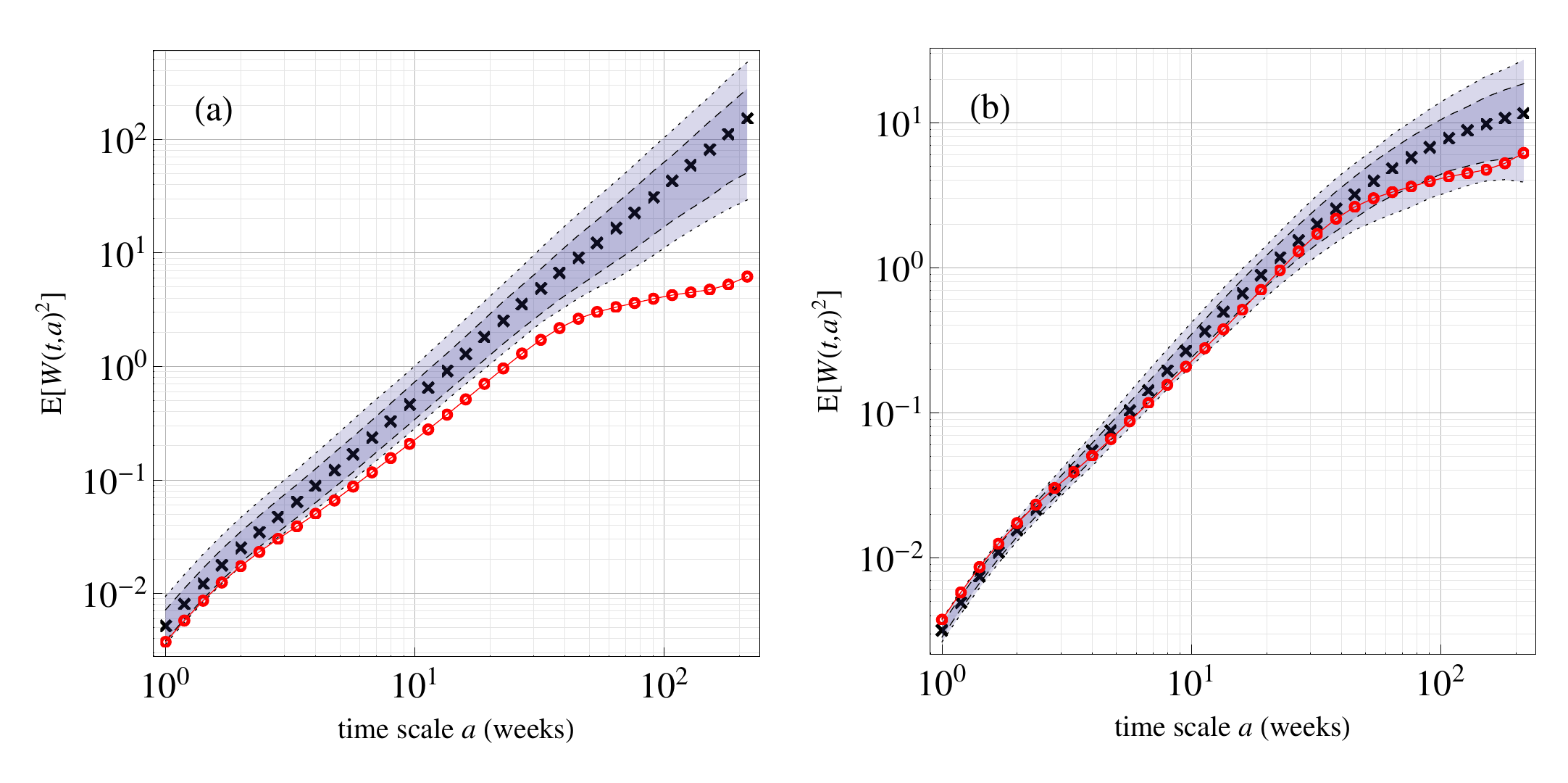}
\caption{(a): Wavelet-based variograms estimated from realizations of the fractional MRW model with parameters $H=0.45$, $\lambda=0.67$, $T =101\, \mbox{weeks}$ and  
$\sigma=0.20$. For each time scale $a$ we have plotted the mean (crosses) and the $1/8$ (dashed) and $1/40$ (dotted) upper and lower quantiles. Prior to the analysis we have added a sinusoidal oscillation with a one-year period and an amplitude $0.25$ to each realization. 
The curve with circles is the wavelet-based variogram for the daily mean (logarithmic) spot price sampled every Monday. (b): The same as (a), but in this case we have simulated the damped MRW model with parameters $1/\nu=21.9\,\mbox{weeks}$,  $\lambda=0.70$, $T=26\,\mbox{weeks}$ and $\sigma= 0.18$. } \label{fig4}
\end{center}
\end{figure}

\section{Maximum likelihood estimators} \label{ML}

\subsection{Computation of the likelihood function for the standard MRW model} \label{MRWML}
The standard MRW model can be written as $x_k = \sigma \sqrt{M_k}\,\varepsilon_k$, where the processes $M_k=c \exp(h_k)$ and $\varepsilon_k$ are as described in section \ref{MRWsection}.   
Given a time series of $n$ observations $z=(z_1,\dots,z_n)$ (which we want to model with the process $x_k$) the ML estimator seeks the parameter vector $\theta=(\lambda,\sigma,T)$ that maximizes the likelihood function $\mathcal{L}_x(\theta|z)$, i.e.  
$$
\hat{\theta} = \op{argmax}_{\theta} \mathcal{L}_x(\theta|z)\,,
$$
where $\mathcal{L}_x(\theta|z) = p_x(z|\theta)$ is the $n$-dimensional probability density function (PDF) for the random vector $x=(x_1,\dots,x_n)$, evaluated at the point $z$ for a fixed parameter vector $\theta$. 

The challenge is to efficiently compute the PDFs $p_x(x)$. Such a method is presented in detail in \cite{Lovsletten:2011vm}, and here we will only explain the main ideas. The first step is to denote $h=(h_1,\dots,h_n)$ and to write 
\begin{equation} \label{ml1}
p_x(x) = \int_{\R^n} p_{x,h}(x,h)\,dh = \int_{\R^n} p_{x|h}(x|h) p_h(h) dh\,,
\end{equation}
where $p_{x,h}$ is the joint PDF for the pair $(x,h) \in \R^n \times \R^n$. Here $p_{x|h}$ is the conditional PDF of $x$ given $h$, and $p_h$ is the marginal PDF of $h$. The first term is easily computed by noting that $x|h$ is a Gaussian vector with independent entries. This gives 
\begin{equation} \label{ml2}
p_{x|h}(x|h) = \prod_{k=1}^n p_{x_k|h_k}(x_k|h_k)\,,
\end{equation} 
where 
\begin{equation} \label{ml3}
p_{x_k|h_k}(x_k|h_k) = \frac{1}{\sqrt{2 \pi\, c \exp(h_k)} \,\sigma } \exp \Big{(} - \frac{x_t^2}{2 \sigma^2 \,c \exp(h_k)} \Big{)}\,.
\end{equation}

By definition the vector $h$ is centered and Gaussian with specified co-variances $\op{Cov}(h_k,h_l)=\gamma(|k-l|)$, and so the second factor $p_h(h)$ is calculated by using standard techniques for Gaussian vectors: For each $k=1,\dots,n$ we define the regression coefficients $\varphi^{(k)}_i$ by the equations
\begin{equation} \label{equations1}
\sum_{j=1}^k \varphi^{(k)}_j \gamma(|i-j|)   = \gamma(i) \mbox{ ~~for~~ } i=1,\dots,k\,.
\end{equation}
Then it holds that 
\begin{equation} \label{ar1}
h_k=\varphi^{(k-1)}_1 h_{k-1} + \dots + \varphi_{k-1}^{(k-1)} h_1 + w_k\,,
\end{equation}
where $w_k$ are independent and centered Gaussian variables with variances 
$$
s^2_k = \gamma(0) - \sum_{i=1}^{k-1} \varphi_i^{(k-1)} \gamma(i)\,.
$$
We can now make an approximation by fixing a truncation parameter $K \in \N$, and for $k>K$ replacing the expression in equation (\ref{ar1}) with 
$$
h_k=\varphi^{(K)}_1 h_{k-1} + \dots + \varphi_{K}^{(K)} h_{k-K} + w_k^{(K)}\,,
$$ 
where $w_k^{(K)}$ are independent and centered Gaussian variables with variances $s_{K+1}^2$. With this approximation we obtain the following expression for $p_h(h)$: 
\begin{eqnarray*}
\log p_h(h) &=& - n \log \sqrt{2 \pi} - \sum_{k=1}^K \log s_K - (n-K) \log s_{K+1} - \sum_{k=1}^K \frac{\big{(}h_k-\varphi^{(k-1)}_1 h_{k-1} - \dots - \varphi_{k-1}^{(k-1)} h_1\big{)}^2}{2 s_k^2} \\
&-&  \sum_{k=K+1}^n \frac{\big{(}h_k-\varphi^{(K)}_1 h_{k-1} - \dots - \varphi_{K}^{(K)} h_{k-K}  \big{)}^2}{2 s_{K+1}^2} \,.
\end{eqnarray*}
By combining this expression with equations (\ref{ml2}) and (\ref{ml3}) we have an expression for the full likelihood $p_{x,h}(x,h)$, and what remains is to calculate the integral in equation (\ref{ml1}).  This integral can be accurately approximated using Laplace's method \cite{Laplace:1986us}. This method entails writing $p_{x,h}(x,h) = \exp(n f_x(h))$ and assuming that the function $f_x(h)$ has a global maximum $h^*$ as a function of $h$. For large $n$ the contribution to the integral in equation (\ref{ml1}) is concentrated around $h^*$, and the hence we can approximate it by making a second order Taylor expansion of $f_x(h)$ about $h^*$. The result is the approximation 
$$
p_x(x) \approx \exp (f_x(h^*)) \int_{\R^n} \exp \Big{(} \frac12 (h-h^*) \Omega_x (h-h^*)^T\Big{)}\,dh = (2 \pi)^{n/2} |\det \Omega_x|^{-1/2}\,p_{x,h}(x,h^*)\,,
$$
where $\Omega_x$ is the Hessian matrix of $f_x(h)$ evaluated at the point $h^*$. 

\subsection{Computation of the likelihood function for the damped MRW model} \label{dampedML}
Let $y_k$ be the damped MRW model defined by equation (\ref{damped}) and $x_k$ be the standard MRW model. We can write 
$$
y_k = \phi y_{k-1} + x_k\,,
$$
and then, for $y=(y_1,\dots,y_n)$ and an initial condition $y_0$, we have
$$
p_y(y) = \int p_{y_0}(y_0) p_x(y_1-\phi y_0,y_2-\phi y_1,\dots,y_{n}-\phi y_{n-1})\,dy_0\,.
$$
For large $n$, the ML method is insensitive to the initial condition $y_0$, and therefore we choose $y_0=0$, which gives   
$$
p_y(y)= p_x(y_1,y_2-\phi y_1,\dots,y_{n}-\phi y_{n-1})\,.
$$
We then have a simple relationship between the likelihood functions of the process $y_k$ and the the likelihood functions for the process $x_k$: 
$$
\mathcal{L}_y\big{(} z_1,\dots,z_n\,|\, (\lambda,\sigma,T,\phi)\big{)} = \mathcal{L}_x\big{(} z_1,z_2-\phi z_1,\dots,z_n-\phi z_{n-1}\,|\, (\lambda,\sigma,T,\phi)\big{)}\,.
$$  

\subsection{Computation of the likelihood function for the fractional MRW model} \label{fractionalML}
In the fractional MRW model the white noise process $\epsilon_k$ is replaced by a fractional Gaussian noise $\varepsilon_k^{(H)}$ with Hurst exponent $H$. By definition this is a centered Gaussian process with co-variance
$$
\beta(|k-l|) \stackrel{\op{def}}{=} \op{Cov}(\varepsilon^{(H)}_k,\varepsilon^{(H)}_l) = \frac{1}{2} \Big{\{} (|k-l|+1)^{2H}- 2 |k-l|^{2H}+(|k-l|-1)^{2H} \Big{\}}\,.
$$  
In the same way as for process $h_k$ in section \ref{MRWML}, we can write 
\begin{equation} \label{ar2}
\varepsilon_k^{(H)} =\xi^{(k-1)}_1  \varepsilon^{(H)}_{k-1} + \dots + \xi_{k-1}^{(k-1)} \varepsilon^{(H)}_1 + w_k\,,
\end{equation}
where the regression coefficients are defined via the equations 
\begin{equation} \label{equations2}
\sum_{j=1}^k \xi^{(k)}_j \beta(|i-j|)   = \beta(i)\, \mbox{,~~ } i=1,\dots,k\,,
\end{equation}
and $w_k$ are independent and centered Gaussian variables with variances 
$$
r^2_k = \beta(0) - \sum_{i=1}^{k-1} \xi_i^{(k-1)} \beta(i)\,.
$$
Again we fix a truncation parameter $K \in \N$, and for $k>K$ we replace the expression in equation (\ref{ar2}) with the expression 
$$
\varepsilon^{(H)}_k=\xi^{(K)}_1 \varepsilon^{(H)}_{k-1} + \dots + \xi_{K}^{(K)} \varepsilon^{(H)}_{k-K} + w_k^{(K)}\,,
$$ 
where $w_k^{(K)}$ are independent and centered Gaussian variables with variances $r_{K+1}^2$. For $k>K$ it follows that the conditional PDF of $x_k$, given both $h_1,\dots,h_k$ and $x_1,\dots,x_{k-1}$, satisfies  
\begin{eqnarray*}
\log p_{x_k|h_1,\dots,h_k,x_1,\dots,x_{k-1}} (x_k|h_1,\dots,h_k,x_1,\dots,x_{k-1}) &=& -\log (\sigma \sqrt{2 \pi c}) - \frac{h_k}{2} - \log  r_{K+1}  \\ 
&-& \frac{1}{2\sigma^2 c r_{K+1}^2} \Bigg{(} x_k \exp(-h_k)- \sum_{i=1}^K \xi^{(K)}_i  x_{k-i} \exp(-h_{k-i}) \Bigg{)}^2\,.
\end{eqnarray*}
From this we obtain an expression for the conditional PDF of $x$ given $h$: 
\begin{equation} \label{cond}
p_{x|h}(x|h) = \sum_{k=1}^n p_{x_k|h_1,\dots,h_k,x_1,\dots,x_{k-1}} (x_k|h_1,\dots,h_k,x_1,\dots,x_{k-1})\,.
\end{equation}
Equation (\ref{cond}) is substituted into equation (\ref{ml1}). The factor $p_h(h)$ and the integral in equation (\ref{ml1}) are computed as explained in section \ref{MRWML}.  

\subsection{Implementation of the ML estimators}
The methods described above are implemented as packages in the R programming language. Equations (\ref{equations1}) and ({\ref{equations2}}) are efficiently solved using the Durbin-Levinson algorithm \cite{Trench:1964vj,McLeod:2007wp}, so the most intensive computations are determining the maxima $h^*$ in the Laplace approximation. This is done using analytic expressions for the derivatives of the functions $f_x(h)$, which roots are found numerically using the derivative-free SANE algorithm \cite{LaCruz:2006vo}. 

The estimates $\hat{\theta}$ are found by numerically optimizing the likelihood functions.

\begin{table}
\begin{center}
\begin{tabular}{||c|c|c|c|c||} \hline
time series & $1/\nu$ (weeks) & $\lambda$ & $T$ (weeks) & $\sigma$ \\ \hline \hline
daily means &29 & 0.68 & 112 & 0.13 \\
daily maxima &22  & 0.86 & 14 & 0.17 \\
 weekly means &40 & 0.55& 132 & 0.14 \\
weekday 1 &22 &  0.70 & 26 & 0.18 \\
weekday 2 &20 & 0.57 & 99 & 0.16 \\
weekday 3 &24 & 0.59 & 99 & 0.16 \\
weekday 4 &26 & 0.64 & 71 & 0.18 \\
weekday 5 &26 & 0.65 & 25 & 0.16 \\
weekday 6 &39 & 0.72 & 32 & 0.17 \\
weekday 7 &41 & 0.78 & 88 & 0.25 \\ \hline \hline
\end{tabular} \caption{ML estimates for the damped MRW model using the method described in section \ref{dampedML}. The first data set consists of the daily mean logarithmic spot price after having subtracted a linear trend. In the second signal we consider daily maxima of the logarithmic price rather than the daily mean price. The third signal is obtained from the first signal my taking seven-day means. The seven remaining signals are obtained from the first signal by sampling every seventh day.} \label{tab1}
\end{center}
\end{table}

\begin{table}[t]
\begin{center}
\begin{tabular}{||c|c|c|c|c||} \hline
time series & $H$ & $\lambda$ & $T$ (weeks) & $\sigma$ \\ \hline \hline
daily means  & 0.47 & 0.68 &17 & 0.10 \\
daily maxima & 0.40 & 0.83 & 16 & 0.16 \\
weekly means & 0.59 & 0.58 & 97 & 0.14 \\
weekday 1 & 0.45 & 0.67& 101 & 0.20 \\
weekday 2 & 0.45 & 0.57& 146 & 0.17 \\
weekday 3 & 0.45 & 0.60 & 99 & 0.17 \\
weekday 4 & 0.44 & 0.64 & 97 & 0.18 \\
weekday 5 & 0.44 & 0.63 & 101 & 0.18 \\
weekday 6 & 0.51 & 0.71 & 73 & 0.19 \\
weekday 7 & 0.50 & 0.80 & 97 & 0.26 \\ \hline \hline
\end{tabular} \caption{ML estimates for the fractional MRW model using the method described in section \ref{fractionalML}. The data sets are as explained in the caption of table \ref{tab1}. } \label{tab2}
\end{center}
\end{table}

\section{Results} \label{results}
The ML estimators for the damped MRW model and the fractional MRW model are applied to the data from the Nord Pool market and the results are shown in tables \ref{tab1} and \ref{tab2}. As discussed in section \ref{data} we use daily averaged data sampled once a week. Hence we have one time series for each day of the week, and these are referred to as ``weekday 1'' to ``weekday 7'' in tables \ref{tab1} and \ref{tab2}.  For comparison we have also included two time series that are sampled daily. These are the daily means and the daily maxima of the logarithmic spot prices. In addition we have looked at the time series consisting of weekly means of the logarithmic prices. 

For the weekly sampled time series (``weekday 1'' to ``weekday 7'') the estimates for the damped MRW model give characteristic time scales $1/\nu$ that vary between 20 and 40 weeks. The mean value is 28 weeks and the standard deviation is 8.2 weeks. These results are consistent with the preliminary analysis discussed in section \ref{data}, where we showed that the wavelet-based variograms fit well with the wavelet-based variogram of an OU process with $1/\nu = 20 \,\mbox{weeks}$. Also, the power spectrum of the spot prices fits with a Lorentzian spectrum with this characteristic time scale.     

The highest estimate of $1/\nu$ are found for the time series ``weekday 7''. This indicates that the anti-correlations are weaker if we consider only the spot prices for Sundays. This is also seen from the estimates for the fractional MRW model. Table \ref{tab2} shows that the estimates of the Hurst exponent is 0.44-0.45 for all the weekdays, but around 0.50 for Saturdays and Sundays. Since the difference between weekdays and weekends primarily is due to the industry's energy demand, this finding indicates that the anti-persistence in electricity spot prices is related to macroeconomic variations rather than mean-reverting mechanisms of natural factors such as weather and water levels.

Relative to what is observed in stock markets we find large values of the intermittency parameter $\lambda$, both for the damped MRW model and for the fractional MRW. For the fractional MRW model the estimates of $\lambda$ vary between the 0.57 and 0.80 for the different weekdays, with an average of $0.66$ and a standard deviation of $0.08$. For the damped MRW model the estimates vary from 0.57 to 0.77, with a mean of 0.66 and a standard deviation of $0.07$. In both cases we observe higher estimates of $\lambda$ for Saturdays and Sundays than for the rest of the week. This could mean that the volatility clustering is stronger for weekend prices than for the rest of the week, but it may also be a simple level-effect. In commodity prices one always observes that the fluctuation level increases with the price itself. This means that when the price is very low, the increments have smaller amplitudes, and this prevents the prices from becoming negative. If the fluctuation level is proportional to the price itself, as in a geometric Brownian motion, then one can normalize for the level effect by taking the logarithm of the price. Although this proportionality hypothesis is a good approximation for most price levels, it might be inaccurate for very low values of the spot price. If this is the case, the fluctuations during times with low prices may be amplified by the logarithmic transformation. This can cause biased estimates of the intermittency parameter and a spurious difference between weekend prices and the rest of the week. 

We finally remark that the estimates of $T$ are known to be very unstable, and it is difficult to draw any meaning from these results.

\section{Comparing the models} \label{testing}
As discussed in sections \ref{introduction} and \ref{data}, the two models presented in this paper, the damped MRW model and the fractional MRW model, represent two different ways of describing the anti-correlations in electricity spot prices. In the damped MRW model the correlations decay exponentially according to equation (\ref{exp}), whereas the correlation decay is a power law, as in equation ({\ref{acf1}), for the fractional model. Our preliminary analysis has revealed that the wavelet-based variograms has a scaling regime up to a time scale of about $20-50$ weeks, and flatten for time scales longer than this. In the same way, the power spectrum flattens for low frequencies. These results can be seen as indications that the damped MRW model is more appropriate than the fractional MRW model for describing spot prices. See figure \ref{prelimfig}(d). On the other hand, the features mentioned above could be effects of the yearly oscillation, since it is known that the existence of periodicities often produce characteristic ``$S$-shapes'' in variograms. We are therefore considering two competing hypotheses. The first hypothesis is that there exists a characteristic scale $1/\nu$ which is not related to the annual periodicity in the signal. More precisely that the correlation decays exponentially with a rate $1/\nu$, and that the non-periodic component of the spot prices can be well described by the damped MRW model. The second hypothesis is that the non-periodic component of the spot price has a algebraic correlation decay on time scales from a day to several years, and that the characteristic scale which is observed is an effect of the annual periodicity. In this case the fractional MRW model is a good description of the spot price data.      

To test the hypotheses we have simulated an ensemble consisting of 500 realizations of the fractional MRW model. The parameters are chosen equal to those estimated for the time series ``weekday 1'', i.e.  $H=0.45$, $\lambda=0.67$, $T =101\, \mbox{weeks}$ and  $\sigma=0.20$. For each of the realizations we have estimated the wavelet-based variogram  
$V(a) = \mathbb{E}[|W(t,a)|^2] $, where $W(t,a)$ is the wavelet transform defined in equation (\ref{wavelet}). For each time scale $a$ the mean is calculated together with the $1/8$ and $1/40$ (upper and lower) quantiles. These curves are plotted in figure \ref{fig4}(a). To simulate the effect of the annual periodicity we have added a sinusoidal oscillation with a one-year period and an amplitude $0.25$. This amplitude is chosen according to the estimate shown in figure \ref{fig1}(d). The analysis shows that the  periodicity is not strong enough to produce the ``break'' in the variograms. Hence this ``break'' represents a characteristic time scale that should be included in the model. In figure \ref{fig4}(b) we show the same analysis as in \ref{fig4}(a), but here we have simulated the damped MRW model with parameters $1/\nu=22\,\mbox{weeks}$,  $\lambda=0.70$, $T=26\,\mbox{weeks}$ and $\sigma= 0.18$. The result shows  that wavelet-based variogram of the spot prices is much better reproduced by the damped model than the fractional model.

\section{Concluding remarks} \label{conclusion}
The main point of this work is to present stochastic models for electricity spot prices that capture both the slowly decaying volatility dependence and the anti-correlated returns. We also present ML methods that efficiently and accurately estimate parameters for these models. For the data from the Nord Pool market the ML estimates show that the intermittency parameter $\lambda$ is significantly higher than what is observed in stock markets, confirming the exceptional nature of electricity spot markets. 

Another important result  is that the damped MRW model performs better than the fractional MRW model, and based on this we conclude that there {\em is} a characteristic scale for the correlation decay of returns (which is not related to the annual oscillations). Estimates show that this time scale is  20-30 weeks. The use of Hurst exponents to characterize correlations is incompatible with the existence of such a characteristic scale, and hence the results of this paper indicate that Hurst-type analysis is inappropriate for electricity spot prices. This conclusion is supported by the fact that the ML estimates of $H$ actually are quite close to $0.5$. If we keep the one-year oscillation in mind, and take into account  that estimates of $H$ tend to be  slightly reduced  in the presence of a periodic component, then it seems unlikely that  $H<0.5$ with any certainty based on the ML estimates. Since the anti-correlations are consistently captured by the damped MRW model, our interpretation is that the fractional model struggles to produce clear evidence for anti-correlations because it imposes a scale free correlation function which is not compatible with the data.  

Another result of this paper becomes evident by considering data sampled on different weekdays. From table $\ref{tab1}$ it appears that the anti-correlations are stronger for weekdays than for weekends. The obvious conclusion from this observation is that the anti-correlation is a result of human activity, and cannot be attributed to natural variations such as weather fluctuations.  

We finally remark that the damped MRW model (toghether with a one-year oscillation) is suitable for forecasting spot prices, and this is the topic of ongoing research. For this purpose, the main advantage of multifractal models is that they efficiently exploit the memory effects in the volatility for forecasts of future prices.

\end{document}